\documentclass[pra,twocolumn]{revtex4}
\usepackage{graphicx}
\usepackage{amsmath}
\usepackage{epsfig}

\begin{document}

\title{Low Distortion Slow Light using Two Absorption Resonances}

\author{Ryan M. Camacho, Michael V. Pack, John C. Howell}

\affiliation{Department of Physics and Astronomy, University of
Rochester, Rochester, NY 14627, USA }

\begin{abstract}
We consider group delay and broadening using two strongly absorbing
and widely spaced resonances.  We derive relations which show that
very large pulse bandwidths coupled with large group delays and
small broadening can be achieved.  Unlike single resonance systems
the dispersive broadening dominates the absorptive broadening which
leads to a dramatic increase in the possible group delay.  We show
that the double resonance systems are excellent candidates for
realizing all-optical delay lines.  We report on an experiment which
achieved up to 50 pulse delays with 40\% broadening.
\end{abstract}

\maketitle \

A  variety of applications in telecommunications and quantum
information have been driving recent interest in slow group
velocities of light pulses. Among these applications are
continuously tunable delay lines, all-optical buffers \cite{boyd05},
optical pattern correlation, ultra-strong cross-phase modulation
\cite{schmidt96}, low light level nonlinear optics
\cite{harris99,lukin00,BEAUSOLEIL04}, and numerous others. The means
for obtaining ultra-slow group velocities have usually involved a
Lorentzian transparency or gain resonance: electro-magnetically
induced transparency (EIT)
\cite{kasapi95,liu01,kash99,Budker99,hau99}, coherent population
oscillations (CPO) \cite{bigelow03,Zhao05,palinginis05}, stimulated
Brillouin scattering (SBS)
\cite{Okawachi05,gonzalez-herraez05,gonzalez-herraez06}, stimulated
Raman scattering (SRS) \cite{dahan05,sharping05} etc..

In this paper we discuss delaying pulses whose center frequency lies
between two strongly absorbing resonances.  Many researchers have
considered using gain doublets in the context of pulse advancement
\cite{chiao93,wang00,dogariu01,macke03,agarwal04}, and  Macke and
Segard \cite{macke03} have discussed pulse advancements for
absorptive doublets.  Grischkowsky \cite{grischkowsky73} measured
the delay of a pulse between two Zeeman-shifted absorbing
resonances, and Tanaka et al. \cite{tanaka03} performed initial
measurements of the delay of a pulse between two atomic hyperfine
resonances.  This work considers both delay and broadening with an
emphasis on the suitability of the delay and broadening
characteristics for practical applications.

In the context of optical delay lines, several criteria must be
satisfied for slow light to be useful. First, the slowed light
pulses must meet system bandwidth specifications.  Second, the
delay-bandwidth product must be much larger than unity. Third, the
delay re-configuration rate should be faster than the inverse pulse
propagation time through the medium. Fourth, pulse absorption must
be kept to a minimum. Fifth, the pulse broadening should also be
minimal. The exact requirements for practical optical buffers are
application dependant. A typical system operating at 10~Gb/sec with
return-to-zero coding and a 50~\% duty cycle might require 7.5~GHz
signal bandwidth, a delay bandwidth product of 1000, a
re-configuration rate in excess of 7.5~MHz with less than 90\%
absorption and pulse broadening of less than 2.

Despite widespread interest in large pulse delays, simultaneously
satisfying all five criteria for most applications has proven
difficult.   In this paper we show that double Lorentzian systems
manages four of these criteria well: large bandwidth, large delay
bandwidth product, minimal absorption and minimal dispersion.
Although we have not realized fast reconfiguration rates, there are
a number of proposals which suggest that fast reconfiguration rates
using double Lorentzians may be possible. In single and double
Lorentzian systems, there exists a tradeoff between large
delay-bandwidth products and pulse broadening. We show that double
Lorentzians in contrast to single Lorentzians have interesting
properties which help minimize this tradeoff while preserving all
other criteria.

Consider two absorbing Lorentzian lines of equal width separated by
a spectral distance much larger than their widths.  Following the
the single Lorentzian formalism of Ref. \cite{boyd02}, the
susceptibility of the double Lorentzian is given by

\begin{equation}
\chi=\beta \left( \frac{1}{\omega_1-\omega-i\gamma}
+\frac{1}{\omega_2-\omega-i\gamma}\right),
\end{equation}
where $\beta$ is the strength of the susceptibility and $2\gamma$ is
the full-width at half-maximum (FWHM).  Making the change of
variables $\omega=({\omega_1+\omega_2})/{2}+\delta$ and
$\omega_0=({\omega_2-\omega_1})/{2}$ and assuming the far detuned
limit (i.e. $\omega_0\gg\gamma$), we may neglect the half-width term
in the denominator. We further assume the pulse frequencies to lie
within the range $|\delta|\ll\omega_0$, the pulse bandwidth to be
larger than the Lorentzian half-width,$\gamma$, and $\chi\ll1$. The
real and imaginary parts of the refractive index
$n=n'+in''\approx1+\chi/2$ may then be written as
\begin{eqnarray}
\nonumber n'&\approx&1+\frac{\beta}{2}
\left(\frac{1}{\delta+\omega_0}+ \frac{1}{\delta-\omega_0}\right) \\
& \approx & 1+\frac{\beta }{\omega_0^2}\delta+\frac{\beta
}{\omega_0^4}\delta^3
\end{eqnarray}
and
\begin{eqnarray}
\nonumber n''&\approx&\frac{\beta \gamma}{2} \left( \frac{1}
{(\delta+\omega_0)^2} + \frac{1} {(\delta-\omega_0)^2}\right)
\\ & \approx& \frac{\beta \gamma}{\omega_0^2}+3\frac{\beta
\gamma}{\omega_0^4}\delta^2,
\end{eqnarray}
where the power series are expanded about $\delta=0$.

The optical depth $\alpha L=2 \omega L n''/c$ (here $L$ is the
interaction length and $\alpha$ is the intensity coefficient) at the
midpoint between the Lorentzians is found to be $\alpha_m L={2
\omega L  \beta \gamma }/{c \omega_0^2}$ which implies ${\partial
n'}/{\partial \delta}|_{\delta=0}={c \alpha_m}/{2 \gamma \omega}$.
The group velocity is then given by
\begin{equation}
v_g\approx \frac{c}{\omega \frac{\partial n'}{\partial
\delta}}=\frac{2\gamma }{\alpha_m},
\end{equation}
and the group delay is given by
\begin{equation}\label{eq:delay}
t_g= \frac{L}{v_g}\approx\frac{\alpha_m L}{2\gamma }.
\end{equation}

The dispersive and absorptive broadenings in the
small-pulse-bandwidth limit (i.e., pulse bandwidth is much smaller
than the spectral distance between Lorentzians) are dominated by the
second terms in the power series expansions of the real and
imaginary parts respectively. The absorptive broadening is due to
the spectrally dependent absorption in the wings of the pulse
spectrum. In the small-pulse-bandwidth limit the absorption can be
approximated by a Gaussian shaped spectral filter plus a constant
absorption:
\begin{equation}
S(\delta)=\exp[-\alpha(\delta)L]\approx\exp[-\alpha_m L-3 \delta^2
\alpha_m L /\omega_0^2]
\end{equation}
When the input pulse is a bandwidth-limited Gaussian, we find that
in the frequency domain the output pulse is the product of the
spectral filter and the input pulse spectrum $A_{in}(\delta)$:
\begin{eqnarray}
\nonumber A_{out}(\delta) &=& A_{in}(\delta)S(\delta)\\
\nonumber &\propto& \exp[-\alpha_m L-\delta^2\left(T_0^2 \ln
2-\frac{3 \alpha_m L}{\omega_0^2}\right)],
\end{eqnarray}
where $T_0$ is the input half-width at half-maximum of the pulse.
Thus, accounting for only absorptive broadening the temporal
half-width after traversing the medium is
\begin{equation}\label{eq:T_a}
T_a=\sqrt{T_0^2+\frac{3 \alpha_m L}{\omega_0^2\ln2}}.
\end{equation}

The temporal broadening due to dispersion is approximated by taking
the difference in group delay for a pulse centered at $\delta = 0$
and a pulse centered at $1/T_0$.  The temporal half-width due to
dispersion is
\begin{equation}\label{eq:T_d}
T_d= T_0 + \frac{3\alpha_m L}{4\ln (2) \gamma \omega_0^2 T_0^2}.
\end{equation}

The total pulse broadening is found by replacing $T_0$ in eq.
\ref{eq:T_d} with $T_a$ from eq. \ref{eq:T_a}:
\begin{eqnarray}\label{eq:T_tot}
T_{tot} = \sqrt{T_0^2+\frac{3 \alpha_m L}{\omega_0^2\ln2}} +
\frac{3\alpha_m L}{4 \gamma \omega_0^2\ln2
\left(T_0^2+\frac{3\alpha_m L}{\omega_0^2\ln2} \right)}.
\end{eqnarray}
We focus on the case where $\omega_0\gg1/T_0\gg\gamma$ and
$T_a/T_0\le2$, corresponding to our experimental parameters. For
this case the dispersive broadening dominates (i.e. the second term
on the right hand side of eq. \ref{eq:T_d} contributes most to the
broadening). However, the quadratic absorption is still significant
since it reduces the effects of dispersive broadening by most
strongly absorbing those frequencies which experience the largest
dispersion (i.e. frequency wings of the pulse). For the parameters
considered in this paper, pulse broadening is less with both
absorptive and dispersive broadening included than for dispersive
broadening alone. In single Lorentzian systems, absorption is the
dominant broadening mechanism and this relationship between
broadening mechanisms is not significant.

Although in hot Rb vapor the resonances experience strong
inhomogeneous Doppler broadening, in the far-wing limit the Rb
resonances are essentially Lorentzian and the double Lorentzian
formalism is a very good approximation. The Rb 85 D$_2$ hyperfine
resonances are separated by approximately 3 GHz, so the gaussian
Doppler broadening of approximately 500 MHz has little effect on the
absorptive behavior. Also, collisional broadening was not
significant for the temperatures used in this work.

\begin{figure}[tb!]
\centerline{\includegraphics[scale=0.45]{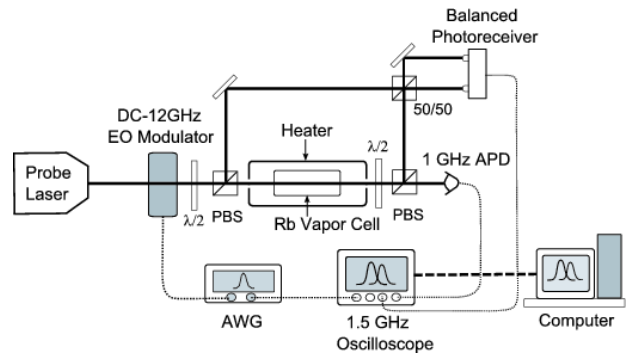}}

\caption{ Experimental schematic.  A probe laser passes through a
heated rubidium vapor cell and is either measuring directly using a
fast detector, or after interference on a balanced photoreceiver.}
\label{experiment}
\end{figure}

A diagram of the experimental setup is shown in Fig.
\ref{experiment}. A narrowband (300 kHz) diode laser at 780 nm
generates a beam of light tuned halfway between the Rb 85 D$_2$
hyperfine resonances, which is fiber coupled into a fast
electro-optic modulator (EOM). An arbitrary waveform generator (AWG)
drives the EOM, producing light pulses with a duration of
approximately 2 ns FWHM.  The light pulses then pass through a 10 cm
glass cell containing rubidium in natural isotopic abundance. The
cell is heated with electronic strip heaters and enclosed in a
cylindrical oven with anti-reflection coated windows.  The pulse is
then incident upon a 1 GHz avalanche photo-diode (APD) and recorded
on a 1.5 GHz oscilloscope triggered by the AWG.

A Mach-Zehnder interferometer was also used with a balanced
photoreciever in order to make continuous wave (CW) measurements of
the transmission and phase delay as a function of frequency. The
difference signal from the balanced photoreciever provides phase
information while transmission data is obtained by blocking one of
the photoreciever photodiodes. The beam splitter preceding the vapor
cell is polarizing to allow for easy balancing of the interferometer
arms, and the beam splitter immediately following the vapor cell is
polarizing to allow switching between the fast APD and CW balanced
detection.

\begin{figure}[tb!]
\centerline{\includegraphics[scale=0.45]{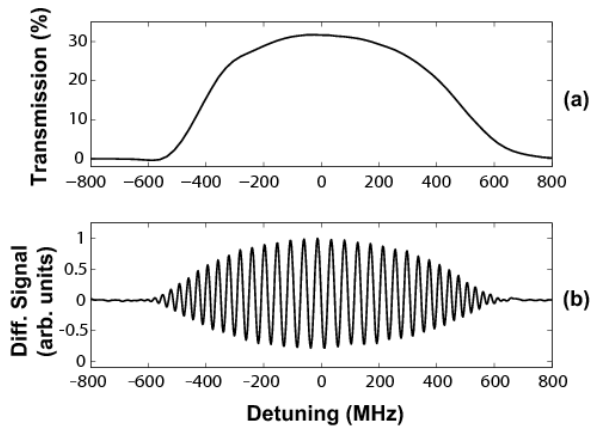}}

\caption{ (a) Probe transmission versus probe detuning and (b)
difference signal from the balanced photoreceiver with each fringe
corresponding to a $2\pi$ phase shift. The height of the fringes is
in arbitrary units. Both transmission and phase data were taken with
a 10 cm Rb vapor cell at approximately 130 C (corresponds to a group
delay of 26~ns)} \label{resonance}
\end{figure}

Figure \ref{resonance} shows (a) absorption  and (b) phase
spectroscopy scans for the transmission window resulting in a
measured 26 ns pulse delay.   The transmission window has a width of
approximately 1 GHz which is sufficient acceptance bandwidth for the
2 ns pulses used in this experiment. The interference fringes were
obtained by sweeping the laser frequency and monitoring the
intensity difference at the two output ports of a Mach-Zehnder
interferometer (see Fig. \ref{experiment}).

It is straightforward to predict the group delay from the absorption
scan or measure it directly using the interference fringes. From the
absorption data, we may extract the optical depth and calculate the
group delay via Eq. 5, giving approximately 26 ns for absorption
data in Fig. \ref{resonance}a, in good agreement with the measured
delay.  In contrast, from the interference fringes we may extract
the group delay directly:
\begin{equation}
t_g= \frac{L}{v_g} = \frac{L\omega \frac{\partial n'}{\partial
\delta}}{c } \approx \frac{\Delta \phi}{\Delta \delta}= \frac{\Delta
N}{\Delta f},
\end{equation}
where $\Delta N$ is the number of fringes in a frequency range
$\Delta f$.  For the resonance shown in Fig. \ref{resonance}b there
are approximately 25 fringes per GHz, giving a predicted optical
delay of ${25}/{1GHz}=$25 ns, also in good agreement with measured
values.  We note that the maximum delay-bandwidth product of a
dispersive medium is approximately given by the maximum number of
interference fringes that can be obtained within the acceptance
bandwidth.

\begin{figure}[tb!]
\centerline{\includegraphics[scale=0.45]{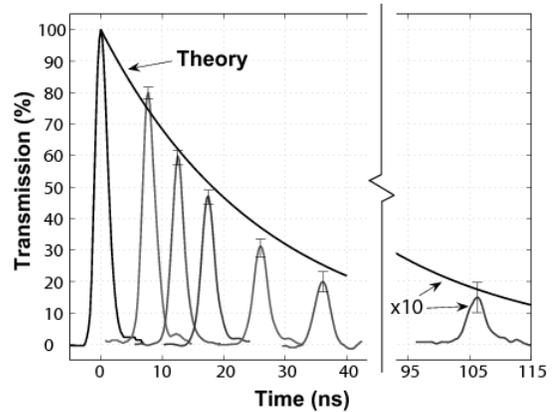}}

\caption{ Pulse delay at various optical depths. On the left, 2.4 ns
pulses are passed through a 10 cm vapor cell and the delay is tuned
by changing temperature. On the right, a 2.1 ns pulse is passed
through four 10 cm cells and delayed 106 ns (50 fractional pulse
delays).} \label{delay}
\end{figure}

Figure \ref{delay} shows probe pulse transmission and delay for
various cell temperatures, plotted in units of percent transmission.
Using a 2.4~ns long pulse (FWHM) and a single 10~cm vapor cell and
varying the temperature between 90 C and 140 C we were able to tune
between 8~ns and 36~ns of delay. We note that several pulse delays
are obtainable with greater than $1/e$ peak transmission. In order
to achieve 106~ns of delay with a delay-bandwidth product of 50 and
a broadening of approximately 40\% we used a 2.1~ns (FWHM) pulse
incident on a series of four 10 cm vapor cells all heated to
approximately 130 C. The theoretical prediction of transmission as a
function of group delay (Eq. 5) is also plotted using the Rb D$_2$
homogeneous linewidth $2\gamma = 2\pi \times 6.07$ MHz from
\cite{Volz96}. The discrepancy between the measured pulse
intensities and the theoretical pulse energies can largely be
attributed to pulse broadening spreading the pulse energy over a
larger time resulting in lower peak intensities.

\begin{figure}[tb!]
\centerline{\includegraphics[scale=0.45]{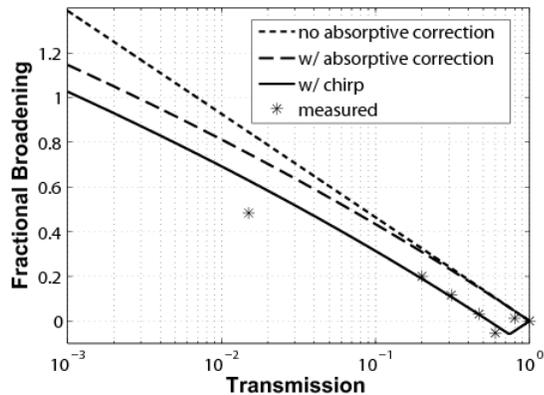}}

\caption{Fractional pulse broadening vs. natural log of
transmission. Fractional broadening is defined as the fractional
increase pulse duration at FWHM (A value of 0 means no broadening).
Due to absorption, the actual broadening is less than that predicted
by the dominant dispersive term, even though absorptive broadening
is negligible. } \label{broadening}
\end{figure}

Figure \ref{broadening} compares the fractional broadening of the
delayed pulses shown in Fig. \ref{delay} to the predicted values
calculated using Eqs. \ref{eq:T_a}-\ref{eq:T_tot}. Shown in Fig.
\ref{broadening} are the measured broadening values, the predicted
total broadening without absorptive corrections,$(T_a+T_d -
T_0)/T_0$, the total predicted broadening, $(T_{tot}-T_0)/T_0$ and
the total predicted broadening with a chirp-like like correction. As
predicted by Eq. 9, the data show that the quadratic absorption
decreases the broadening due to dispersion. Also, for small optical
depths the pulse width compresses before broadening, which may be
modeled by assuming a small negative chirp on the input pulse. We do
not know the origins of the chirp, but we found that by including a
small second order chirp in the theory we obtained a very good fit
to the data.

In conclusion, we have discussed the delay and broadening
characteristics for pulses propagating through a double-Lorentzian
medium (i.e. a medium with two widely spaced absorbing Lorentzian
resonances). For many slow-light applications, absorptive double
Lorentzian systems seem to be better suited than gain-like single
Lorentzian systems. Since the spacing between the two Lorentzians
can be arbitrarily large, the usable bandwidth may be
proportionately large, though practical considerations may limit the
separation. Also, in contrast to single Lorenztians, the
double-Lorentzian lineshape is dominated by dispersive broadening
and not absorptive broadening, resulting in less pulse distortion
for a given delay.  While the method of tuning the delay in the
present experiment was slow (increasing the temperature of the vapor
cell), there may be ways to to achieve fast reconfiguration rates.
Some possibilities may be to drive a large number of atoms to
saturation with a strong auxiliary beam, or make use of light
induced desorption \cite{alexadrov02} of Rb to optically change the
atomic number density.

This work was supported by DARPA Slow Light, the National Science
Foundation, and Research Corporation.
\bibliography{slow1}

\begin{thebibliography}{28}
\expandafter\ifx\csname natexlab\endcsname\relax\def\natexlab#1{#1}\fi
\expandafter\ifx\csname bibnamefont\endcsname\relax
  \def\bibnamefont#1{#1}\fi
\expandafter\ifx\csname bibfnamefont\endcsname\relax
  \def\bibfnamefont#1{#1}\fi
\expandafter\ifx\csname citenamefont\endcsname\relax
  \def\citenamefont#1{#1}\fi
\expandafter\ifx\csname url\endcsname\relax
  \def\url#1{\texttt{#1}}\fi
\expandafter\ifx\csname urlprefix\endcsname\relax\def\urlprefix{URL }\fi
\providecommand{\bibinfo}[2]{#2}
\providecommand{\eprint}[2][]{\url{#2}}

\bibitem[{\citenamefont{Boyd et~al.}(2005)\citenamefont{Boyd, Gauthier, Gaeta,
  and Willner}}]{boyd05}
\bibinfo{author}{\bibfnamefont{R.~W.} \bibnamefont{Boyd}},
  \bibinfo{author}{\bibfnamefont{D.~J.} \bibnamefont{Gauthier}},
  \bibinfo{author}{\bibfnamefont{A.~L.} \bibnamefont{Gaeta}}, \bibnamefont{and}
  \bibinfo{author}{\bibfnamefont{A.~E.} \bibnamefont{Willner}},
  \bibinfo{journal}{Phys. Rev. A} \textbf{\bibinfo{volume}{71}},
  \bibinfo{pages}{023801} (\bibinfo{year}{2005}).

\bibitem[{\citenamefont{Schmidt and Imamo¡glu}(1996)}]{schmidt96}
\bibinfo{author}{\bibfnamefont{H.}~\bibnamefont{Schmidt}} \bibnamefont{and}
  \bibinfo{author}{\bibfnamefont{A.}~\bibnamefont{Imamo¡glu}},
  \bibinfo{journal}{Optics Letters} \textbf{\bibinfo{volume}{21}},
  \bibinfo{pages}{1936} (\bibinfo{year}{1996}).

\bibitem[{\citenamefont{Harris and Hau}(1999)}]{harris99}
\bibinfo{author}{\bibfnamefont{S.~E.} \bibnamefont{Harris}} \bibnamefont{and}
  \bibinfo{author}{\bibfnamefont{L.~V.} \bibnamefont{Hau}},
  \bibinfo{journal}{Phys. Rev. Lett.} \textbf{\bibinfo{volume}{82}},
  \bibinfo{pages}{4611} (\bibinfo{year}{1999}).

\bibitem[{\citenamefont{Lukin and Imamoglu}(2000)}]{lukin00}
\bibinfo{author}{\bibfnamefont{M.~D.} \bibnamefont{Lukin}} \bibnamefont{and}
  \bibinfo{author}{\bibfnamefont{A.}~\bibnamefont{Imamoglu}},
  \bibinfo{journal}{Phys. Rev. Lett.} \textbf{\bibinfo{volume}{84}},
  \bibinfo{pages}{1419} (\bibinfo{year}{2000}).

\bibitem[{\citenamefont{Beausoleil et~al.}(2004)\citenamefont{Beausoleil,
  Munro, and Spiller}}]{BEAUSOLEIL04}
\bibinfo{author}{\bibfnamefont{R.~G.} \bibnamefont{Beausoleil}},
  \bibinfo{author}{\bibfnamefont{W.~J.} \bibnamefont{Munro}}, \bibnamefont{and}
  \bibinfo{author}{\bibfnamefont{T.~P.} \bibnamefont{Spiller}},
  \bibinfo{journal}{J. Mod. Opt.} \textbf{\bibinfo{volume}{51}},
  \bibinfo{pages}{1559} (\bibinfo{year}{2004}).

\bibitem[{\citenamefont{Kasapi et~al.}(1995)\citenamefont{Kasapi, Jain, Yin,
  and Harris}}]{kasapi95}
\bibinfo{author}{\bibfnamefont{A.}~\bibnamefont{Kasapi}},
  \bibinfo{author}{\bibfnamefont{M.}~\bibnamefont{Jain}},
  \bibinfo{author}{\bibfnamefont{G.}~\bibnamefont{Yin}}, \bibnamefont{and}
  \bibinfo{author}{\bibfnamefont{S.}~\bibnamefont{Harris}},
  \bibinfo{journal}{Phys. Rev. Lett.} \textbf{\bibinfo{volume}{74}},
  \bibinfo{pages}{2447} (\bibinfo{year}{1995}).

\bibitem[{\citenamefont{Liu et~al.}(2001)\citenamefont{Liu, Dutton, Behroozi,
  and Hau}}]{liu01}
\bibinfo{author}{\bibfnamefont{C.}~\bibnamefont{Liu}},
  \bibinfo{author}{\bibfnamefont{Z.}~\bibnamefont{Dutton}},
  \bibinfo{author}{\bibfnamefont{C.~H.} \bibnamefont{Behroozi}},
  \bibnamefont{and} \bibinfo{author}{\bibfnamefont{L.~V.} \bibnamefont{Hau}},
  \bibinfo{journal}{Nature} \textbf{\bibinfo{volume}{409}},
  \bibinfo{pages}{490} (\bibinfo{year}{2001}).

\bibitem[{\citenamefont{Kash et~al.}(1999)\citenamefont{Kash, Sautenkov,
  Zibrov, Hollberg, Welch, Lukin, Rostovtsev, Fry, and Scully}}]{kash99}
\bibinfo{author}{\bibfnamefont{M.~M.} \bibnamefont{Kash}},
  \bibinfo{author}{\bibfnamefont{V.~A.} \bibnamefont{Sautenkov}},
  \bibinfo{author}{\bibfnamefont{A.~S.} \bibnamefont{Zibrov}},
  \bibinfo{author}{\bibfnamefont{L.}~\bibnamefont{Hollberg}},
  \bibinfo{author}{\bibfnamefont{G.~R.} \bibnamefont{Welch}},
  \bibinfo{author}{\bibfnamefont{M.~D.} \bibnamefont{Lukin}},
  \bibinfo{author}{\bibfnamefont{Y.}~\bibnamefont{Rostovtsev}},
  \bibinfo{author}{\bibfnamefont{E.~S.} \bibnamefont{Fry}}, \bibnamefont{and}
  \bibinfo{author}{\bibfnamefont{M.~O.} \bibnamefont{Scully}},
  \bibinfo{journal}{Phys. Rev. Lett.} \textbf{\bibinfo{volume}{82}},
  \bibinfo{pages}{5229} (\bibinfo{year}{1999}).

\bibitem[{\citenamefont{Budker et~al.}(1999)\citenamefont{Budker, Kimball,
  Rochester, and Yashchuk}}]{Budker99}
\bibinfo{author}{\bibfnamefont{D.}~\bibnamefont{Budker}},
  \bibinfo{author}{\bibfnamefont{D.~F.} \bibnamefont{Kimball}},
  \bibinfo{author}{\bibfnamefont{S.~M.} \bibnamefont{Rochester}},
  \bibnamefont{and} \bibinfo{author}{\bibfnamefont{V.~V.}
  \bibnamefont{Yashchuk}}, \bibinfo{journal}{Phys. Rev. Lett.}
  \textbf{\bibinfo{volume}{83}}, \bibinfo{pages}{1767} (\bibinfo{year}{1999}).

\bibitem[{\citenamefont{Hau et~al.}(1999)\citenamefont{Hau, Harris, Dutton, and
  Behroozi}}]{hau99}
\bibinfo{author}{\bibfnamefont{L.~V.} \bibnamefont{Hau}},
  \bibinfo{author}{\bibfnamefont{S.~E.} \bibnamefont{Harris}},
  \bibinfo{author}{\bibfnamefont{Z.}~\bibnamefont{Dutton}}, \bibnamefont{and}
  \bibinfo{author}{\bibfnamefont{C.}~\bibnamefont{Behroozi}},
  \bibinfo{journal}{Nature} \textbf{\bibinfo{volume}{397}},
  \bibinfo{pages}{594} (\bibinfo{year}{1999}).

\bibitem[{\citenamefont{Bigelow et~al.}(2003)\citenamefont{Bigelow, Lepeshkin,
  and Boyd}}]{bigelow03}
\bibinfo{author}{\bibfnamefont{M.~S.} \bibnamefont{Bigelow}},
  \bibinfo{author}{\bibfnamefont{N.~N.} \bibnamefont{Lepeshkin}},
  \bibnamefont{and} \bibinfo{author}{\bibfnamefont{R.~W.} \bibnamefont{Boyd}},
  \bibinfo{journal}{Science} \textbf{\bibinfo{volume}{301}},
  \bibinfo{pages}{200} (\bibinfo{year}{2003}).

\bibitem[{\citenamefont{Zhao et~al.}(2005)\citenamefont{Zhao, Palinginis,
  Pesala, Chang-Hasnain, and Hemmer}}]{Zhao05}
\bibinfo{author}{\bibfnamefont{X.}~\bibnamefont{Zhao}},
  \bibinfo{author}{\bibfnamefont{P.}~\bibnamefont{Palinginis}},
  \bibinfo{author}{\bibfnamefont{B.}~\bibnamefont{Pesala}},
  \bibinfo{author}{\bibfnamefont{C.}~\bibnamefont{Chang-Hasnain}},
  \bibnamefont{and} \bibinfo{author}{\bibfnamefont{P.}~\bibnamefont{Hemmer}},
  \bibinfo{journal}{Optics Express} \textbf{\bibinfo{volume}{93}},
  \bibinfo{pages}{7899} (\bibinfo{year}{2005}).

\bibitem[{\citenamefont{Palinginis et~al.}(2005)\citenamefont{Palinginis,
  Sedgwick, Crankshaw, Moewe, and Chang-Hasnain}}]{palinginis05}
\bibinfo{author}{\bibfnamefont{P.}~\bibnamefont{Palinginis}},
  \bibinfo{author}{\bibfnamefont{F.}~\bibnamefont{Sedgwick}},
  \bibinfo{author}{\bibfnamefont{S.}~\bibnamefont{Crankshaw}},
  \bibinfo{author}{\bibfnamefont{M.}~\bibnamefont{Moewe}}, \bibnamefont{and}
  \bibinfo{author}{\bibfnamefont{C.~J.} \bibnamefont{Chang-Hasnain}},
  \bibinfo{journal}{Optics Express} \textbf{\bibinfo{volume}{13}},
  \bibinfo{pages}{9909} (\bibinfo{year}{2005}).

\bibitem[{\citenamefont{Okawachi et~al.}(2005)\citenamefont{Okawachi, Bigelow,
  Sharping, Zhu, Schweinsberg, Gauthier, Boyd, and Gaeta}}]{Okawachi05}
\bibinfo{author}{\bibfnamefont{Y.}~\bibnamefont{Okawachi}},
  \bibinfo{author}{\bibfnamefont{M.~S.} \bibnamefont{Bigelow}},
  \bibinfo{author}{\bibfnamefont{J.~E.} \bibnamefont{Sharping}},
  \bibinfo{author}{\bibfnamefont{Z.}~\bibnamefont{Zhu}},
  \bibinfo{author}{\bibfnamefont{A.}~\bibnamefont{Schweinsberg}},
  \bibinfo{author}{\bibfnamefont{D.~J.} \bibnamefont{Gauthier}},
  \bibinfo{author}{\bibfnamefont{R.~W.} \bibnamefont{Boyd}}, \bibnamefont{and}
  \bibinfo{author}{\bibfnamefont{A.~L.} \bibnamefont{Gaeta}},
  \bibinfo{journal}{Phys. Rev. Lett.} \textbf{\bibinfo{volume}{94}},
  \bibinfo{pages}{153902} (\bibinfo{year}{2005}).

\bibitem[{\citenamefont{González-Herráez
  et~al.}(2005)\citenamefont{González-Herráez, Song, and
  Thévenaz}}]{gonzalez-herraez05}
\bibinfo{author}{\bibfnamefont{M.}~\bibnamefont{González-Herráez}},
  \bibinfo{author}{\bibfnamefont{K.-Y.} \bibnamefont{Song}}, \bibnamefont{and}
  \bibinfo{author}{\bibfnamefont{L.}~\bibnamefont{Thévenaz}},
  \bibinfo{journal}{Appl. Phys. Lett.} \textbf{\bibinfo{volume}{87}},
  \bibinfo{pages}{081113} (\bibinfo{year}{2005}).

\bibitem[{\citenamefont{González-Herráez
  et~al.}(2006)\citenamefont{González-Herráez, Song, and
  Thévenaz}}]{gonzalez-herraez06}
\bibinfo{author}{\bibfnamefont{M.}~\bibnamefont{González-Herráez}},
  \bibinfo{author}{\bibfnamefont{K.-Y.} \bibnamefont{Song}}, \bibnamefont{and}
  \bibinfo{author}{\bibfnamefont{L.}~\bibnamefont{Thévenaz}},
  \bibinfo{journal}{Optics Express} \textbf{\bibinfo{volume}{14}},
  \bibinfo{pages}{1400} (\bibinfo{year}{2006}).

\bibitem[{\citenamefont{Dahan and Eisenstein}(2005)}]{dahan05}
\bibinfo{author}{\bibfnamefont{D.}~\bibnamefont{Dahan}} \bibnamefont{and}
  \bibinfo{author}{\bibfnamefont{G.}~\bibnamefont{Eisenstein}},
  \bibinfo{journal}{Optics Express} \textbf{\bibinfo{volume}{13}},
  \bibinfo{pages}{6234} (\bibinfo{year}{2005}).

\bibitem[{\citenamefont{Sharping et~al.}(2005)\citenamefont{Sharping, Okawachi,
  and Gaeta}}]{sharping05}
\bibinfo{author}{\bibfnamefont{J.~E.} \bibnamefont{Sharping}},
  \bibinfo{author}{\bibfnamefont{Y.}~\bibnamefont{Okawachi}}, \bibnamefont{and}
  \bibinfo{author}{\bibfnamefont{A.~L.} \bibnamefont{Gaeta}},
  \bibinfo{journal}{Optics Express} \textbf{\bibinfo{volume}{13}},
  \bibinfo{pages}{6092} (\bibinfo{year}{2005}).

\bibitem[{\citenamefont{Chiao}(1993)}]{chiao93}
\bibinfo{author}{\bibfnamefont{R.~Y.} \bibnamefont{Chiao}},
  \bibinfo{journal}{Phys. Rev. A} \textbf{\bibinfo{volume}{48}},
  \bibinfo{pages}{R34} (\bibinfo{year}{1993}).

\bibitem[{\citenamefont{Wang et~al.}(2000)\citenamefont{Wang, Kuzmich, and
  Dogariu}}]{wang00}
\bibinfo{author}{\bibfnamefont{L.~J.} \bibnamefont{Wang}},
  \bibinfo{author}{\bibfnamefont{A.}~\bibnamefont{Kuzmich}}, \bibnamefont{and}
  \bibinfo{author}{\bibfnamefont{A.}~\bibnamefont{Dogariu}},
  \bibinfo{journal}{Nature} \textbf{\bibinfo{volume}{406}},
  \bibinfo{pages}{277} (\bibinfo{year}{2000}).

\bibitem[{\citenamefont{Dogariu et~al.}(2001)\citenamefont{Dogariu, Kuzmich,
  and Wang}}]{dogariu01}
\bibinfo{author}{\bibfnamefont{A.}~\bibnamefont{Dogariu}},
  \bibinfo{author}{\bibfnamefont{A.}~\bibnamefont{Kuzmich}}, \bibnamefont{and}
  \bibinfo{author}{\bibfnamefont{L.~J.} \bibnamefont{Wang}},
  \bibinfo{journal}{Phys. Rev. A} \textbf{\bibinfo{volume}{63}},
  \bibinfo{pages}{053806} (\bibinfo{year}{2001}).

\bibitem[{\citenamefont{Macke and Segard}(2003)}]{macke03}
\bibinfo{author}{\bibfnamefont{B.}~\bibnamefont{Macke}} \bibnamefont{and}
  \bibinfo{author}{\bibfnamefont{B.}~\bibnamefont{Segard}},
  \bibinfo{journal}{Eur. Phys. J. D} \textbf{\bibinfo{volume}{23}},
  \bibinfo{pages}{125} (\bibinfo{year}{2003}).

\bibitem[{\citenamefont{Agarwal and Dasgupta}(2004)}]{agarwal04}
\bibinfo{author}{\bibfnamefont{G.~S.} \bibnamefont{Agarwal}} \bibnamefont{and}
  \bibinfo{author}{\bibfnamefont{S.}~\bibnamefont{Dasgupta}},
  \bibinfo{journal}{Phys. Rev. A} \textbf{\bibinfo{volume}{70}},
  \bibinfo{pages}{023802} (\bibinfo{year}{2004}).

\bibitem[{\citenamefont{Grischkowsky}(1973)}]{grischkowsky73}
\bibinfo{author}{\bibfnamefont{D.}~\bibnamefont{Grischkowsky}},
  \bibinfo{journal}{Phys. Rev. A} \textbf{\bibinfo{volume}{7}},
  \bibinfo{pages}{2096} (\bibinfo{year}{1973}).

\bibitem[{\citenamefont{Tanaka et~al.}(2003)\citenamefont{Tanaka, Niwa, Hayami,
  Furue, Nakayama, Kohmoto, Kunitomo, and Fukuda}}]{tanaka03}
\bibinfo{author}{\bibfnamefont{H.}~\bibnamefont{Tanaka}},
  \bibinfo{author}{\bibfnamefont{H.}~\bibnamefont{Niwa}},
  \bibinfo{author}{\bibfnamefont{K.}~\bibnamefont{Hayami}},
  \bibinfo{author}{\bibfnamefont{S.}~\bibnamefont{Furue}},
  \bibinfo{author}{\bibfnamefont{K.}~\bibnamefont{Nakayama}},
  \bibinfo{author}{\bibfnamefont{T.}~\bibnamefont{Kohmoto}},
  \bibinfo{author}{\bibfnamefont{M.}~\bibnamefont{Kunitomo}}, \bibnamefont{and}
  \bibinfo{author}{\bibfnamefont{Y.}~\bibnamefont{Fukuda}},
  \bibinfo{journal}{Phys. Rev. A} \textbf{\bibinfo{volume}{68}},
  \bibinfo{pages}{053801} (\bibinfo{year}{2003}).

\bibitem[{\citenamefont{Boyd and Gauthier}(2002)}]{boyd02}
\bibinfo{author}{\bibfnamefont{R.}~\bibnamefont{Boyd}} \bibnamefont{and}
  \bibinfo{author}{\bibfnamefont{D.~J.} \bibnamefont{Gauthier}},
  \emph{\bibinfo{title}{Progress in Optics}} (\bibinfo{publisher}{Elsevier},
  \bibinfo{year}{2002}), chap. \bibinfo{chapter}{Slow and Fast Light}, p.
  \bibinfo{pages}{497}.

\bibitem[{\citenamefont{Volz and Schmoranzer}(1996)}]{Volz96}
\bibinfo{author}{\bibfnamefont{U.}~\bibnamefont{Volz}} \bibnamefont{and}
  \bibinfo{author}{\bibfnamefont{H.}~\bibnamefont{Schmoranzer}},
  \bibinfo{journal}{Physica Scripta T65} p.~\bibinfo{pages}{48}
  (\bibinfo{year}{1996}).

\bibitem[{\citenamefont{Alexandrov et~al.}(2002)\citenamefont{Alexandrov,
  Balabas, Budker, English, Kimball, Li, and Yashchuk}}]{alexadrov02}
\bibinfo{author}{\bibfnamefont{E.~B.} \bibnamefont{Alexandrov}},
  \bibinfo{author}{\bibfnamefont{M.~V.} \bibnamefont{Balabas}},
  \bibinfo{author}{\bibfnamefont{D.}~\bibnamefont{Budker}},
  \bibinfo{author}{\bibfnamefont{D.}~\bibnamefont{English}},
  \bibinfo{author}{\bibfnamefont{D.~F.} \bibnamefont{Kimball}},
  \bibinfo{author}{\bibfnamefont{C.}~\bibnamefont{Li}}, \bibnamefont{and}
  \bibinfo{author}{\bibfnamefont{V.~V.} \bibnamefont{Yashchuk}},
  \bibinfo{journal}{Phys. Rev. A} \textbf{\bibinfo{volume}{66}},
  \bibinfo{pages}{042903} (\bibinfo{year}{2002}).

\end{thebibliography}

\end{document}